# Spin glass experiments

Eric VINCENT, SPEC, CEA, CNRS, Université Paris-Saclay, 91191, Gif-sur-Yvette, France

## Abstract

A spin glass is a diluted magnetic material in which the magnetic moments are randomly interacting, with a huge number of metastable states which prevent reaching equilibrium. Spin-glass models are conceptually simple, but require very sophisticated treatments. These models have become a paradigm for the understanding of glassy materials and also for the solution of complex optimization problems. After cooling from the paramagnetic phase, the spin glass remains out of equilibrium, and slowly evolves. This aging phenomenon corresponds to the growth of a mysterious "spin-glass order", whose correlation length can be measured. A cooling temperature step during aging causes a partial "rejuvenation", while the "memory" of previous aging is stored and can be retrieved. Many glassy materials present aging, and rejuvenation and memory effects can be found in some cases, but they are usually less pronounced. Numerical simulations of these phenomena are presently under active development using custom-built supercomputers. A general understanding of the glassy systems, for which spin glasses bring a prominent insight, is still under construction.

## 1. Introduction

A spin glass is a set of interacting magnetic moments, originating from spins, in which the magnetic interactions are randomly distributed in sign. In the well-known case of ferromagnets, the interactions are positive, tending to align the moments along the same direction, and producing a net macroscopic magnetization. In antiferromagnets, the interactions are negative, driving the moments to anti-alignment, and establishing a set of two interpenetrating ferromagnetic sublattices, oriented in opposite directions.

We can simply describe the case of spin glasses as a mixture of both ferro- and antiferromagnetic situations. The theoretical definition of the spin glass is that of an ensemble of randomly interacting magnetic moments. The total energy $H$ is the sum over interacting neighbours ($S_i$, $S_j$) of the coupling energies $J_{i,j}S_iS_j$, where the $\{J_{i,j}\}$ are random variables, usually gaussian or $\pm J$ distributed (Edwards and Anderson, 1975) :

$$H = -\sum_{i,j} J_{i,j} S_i S_j \qquad (1).$$

The magnetic moments (or spins) are in random sign interactions, that is, each moment experiences contradicting constraints from its neighbours. This situation of contradicting influences has been termed *frustration*. No simple symmetric configuration of the set of spins corresponds to an equilibrium state. Conversely, the abundance of different spin configurations which have similar energy values produces a huge number of metastable states. The states are separated by free-energy barriers of all heights, which generate a very wide spectrum of response times, from paramagnetic times ($\approx 10^{-12}$ s) up to astronomical scales. Finding the absolute minimum is thus extremely difficult and, from a practical point of view, a spin glass is virtually always out of equilibrium. At low temperature, the spin glass is in a (partially) frozen phase, but this state slowly evolves, with no apparent time limit (for a comprehensive presentation of spin glasses and other disordered magnetic systems, see Nordblad, 2023).

In a spin glass, the disorder lies in the magnetic interactions, which are fixed. Contrary to this situation of a so-called *frozen* disorder, the disorder in usual (structural) glasses lies in the random positions of the molecules. A structural glass is a frozen liquid, in which the molecules are slowly moving, thereby changing the disorder configuration. The spin glass problem, with frozen disorder, is conceptually simpler. It has allowed rich, far-reaching theoretical developments : mean-field treatment by Sherrington and Kirkpatrick, 1975, with a solution by Parisi in a model with "continuous RSB", standing for *continuous replica symmetry breaking*, a sophisticated mathematical method (Parisi, 1979; Mezard et al, 1987; Altieri and Baity-Jesi, 2023). The numerical simulations have brought many results, up to spectacular recent progresses by the Janus collaboration with special-purpose supercomputers (Baity-Jesi et al, 2018; Zhai et al, 2020; Baity-Jesi et al, 2022). Although involving different natures of disorder, both spin glasses and structural glasses share a lot of similitudes, and the spin glass is usually considered a powerful model for the description and understanding of various glassy materials, among which "ferroic" materials have recently received a particular interest (Lookman and Ren, 2018; Ji et al, 2023).

The first studied spin-glass materials were non-magnetic metals (Au, Ag, Pt…) in which a few percent of magnetic atoms (Fe, Mn…) were dispersed at random (Cannella and Mydosh, 1972). The magnetic atoms are separated by random distances, and the oscillating character of the Ruderman-Kittel-Kasuya-Yosida (RKKY) interaction with respect to distance makes their coupling energies take a random sign. As explained in (Mydosh, 1993), the term "spin glass" was first suggested by B.R. Coles in 1970 to be applied to the strange magnetic behaviour of an Au:Co alloy, and appeared again at Cole's instigation in (Anderson, 1970). Examples of spin glasses have also been found within diluted magnetic insulators with shorter-ranged interactions (Alba et al, 1982). Although chemically very different, these various compounds have been found to show a common general behaviour that is now understood as generic for spin glasses. A fairly comprehensive review of the theoretical and experimental aspects of spin glasses can be found in Kawamura and Taniguchi (2015). Another interesting review has been given by J.A. Mydosh (2015), with emphasis on new spin glass materials and related topical problems in condensed matter physics.

# 2. The spin-glass phase

The spin-glass phase forms below a characteristic glass temperature $T_g$, whose magnitude is governed by the strength of the interactions. At temperatures $T$ much higher than $T_g$, thermal magnetic fluctuations dominate against interactions, and the spins are fluctuating paramagnetically (Fig. 1a). The magnetization $M$ obtained under a low field $H$ follows a Curie-Weiss law

$$M/H \propto C/(T-\theta) \qquad (2)$$

which is characteristic of a paramagnetic phase ($C$ is proportional to the square of the individual magnetic moments, and $\theta$ is a temperature proportional to the energy of the interactions).

## 2.1 A frozen state

Below $T_g$, the magnetic behaviour as a function of temperature becomes history-dependent, the spins are in a frozen phase. In the "FC" (field-cooled) procedure, a magnetic field is applied above $T_g$, and *M(T)* is measured upon slowly cooling in the field. The same *M(T)* curve is obtained upon re-heating, the FC curve is usually stable within 1%, and is mostly flat.

On the other hand, in the "ZFC" (zero-field cooled) procedure, the sample is cooled in zero field, and the field is applied at the lowest temperature. Since the spins are in a frozen state, the magnetization cannot reach an equilibrium value (which we consider to be the FC value). It slowly increases with time. Then, as we increase the temperature, the magnetization curve *M(T)* progressively rises up, finally merging with the FC curve in the vicinity of $T_g$.

Fig. 1a shows typical FC-ZFC measurements (Dupuis, 2002). When cooling from the paramagnetic phase, the magnetization rises up, until reaching $T_g$, below which the FC curve is essentially flat. This apparent "solidification" of the spin system at $T_g$, with no further significant evolution of the magnetization, suggests a collective behaviour of the spins, with a sudden freezing into a long-range correlated state. We show below that the process of freezing at $T_g$ presents a critical character, and we describe how this amazing *glassy ordered* state of matter can be characterized.

We can also explore the spin-glass phase by means of *ac* field measurements, which usefully complete the *dc* procedures described above. When a weak *ac* magnetic field is applied, the observed *ac* susceptibility $\chi$ is usually decomposed into two components : $\chi'$ (in-phase component), and $\chi''$ (out-of-phase, or dissipative, component). Fig. 1b shows the *ac* susceptibility of a spin glass at frequency $\omega/2\pi$ = 0.4Hz, as a function of temperature (Dupuis, 2002). $\chi'$ has the same shape as the ZFC (*dc*) magnetization in Fig. 1a, which can be thought as an *ac* measurement at a frequency that is the inverse of the time spent at each heating temperature step in the ZFC curve.

The out-of-phase response $\chi''$ is null above $T_g$, in the paramagnetic phase the spins are oscillating in phase with the excitation field. Upon cooling down, the approach of $T_g$ is signalled by a sharp increase of $\chi''$ : the magnetic response is delayed, the spin system is freezing.

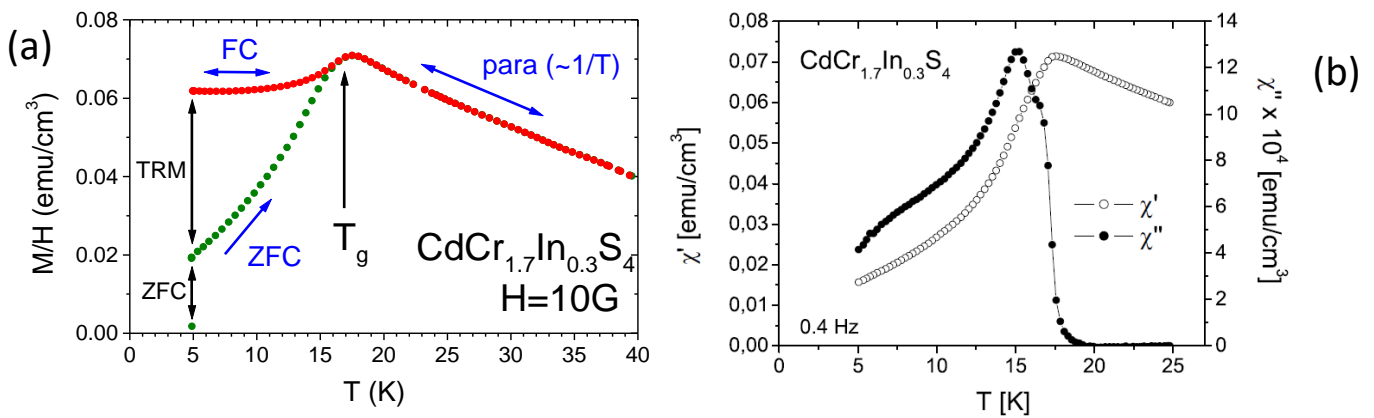

**Figure 1** Typical magnetic characterization of a spin glass (Dupuis, 2002). The sample is an insulating spin glass prepared by M. Noguès (Alba et al, 1982). **Left (a) :** *dc* measurements. Above the glass temperature $T_g$=16.7K, the magnetization has a paramagnetic shape. In the spin-glass phase, below $T_g$, the field-cooled (FC) and zero-field cooled (ZFC) curves are split, showing a history-dependent behaviour. **Right (b) :** *ac* measurements. The in-phase component $\chi'$ of the *ac* susceptibility has a peak at a frequency-dependent freezing temperature. The out-of-phase (dissipative) component $\chi''$ is null in the paramagnetic phase, and rises up when entering the spin-glass phase.

## 2.2 Critical dynamics of freezing

The $\chi'$ peak coincides with the inflection point of the $\chi''$ rise. It takes place at a freezing temperature $T_f(\omega)$ which depends on the frequency $\omega/2\pi$ : for higher frequencies, higher freezing temperatures are observed, in the same way as, on a photo taken with a shorter aperture time, faster events can be seen being still. We consider that $T_g$ is the zero-frequency limit of $T_f(\omega)$. Studying the frequency dependence of $T_f(\omega)$ allows checking whether the freezing transition at $T_g$ is a dynamical process (like for independent superparamagnetic particles), or presents the collective character of a thermodynamic phase transition (Tholence, 1981).

If the freezing process at $T_g$ is a collective process, then there should be a divergence of a correlation length $\xi$ at the approach of the transition at $T_g$, according to :

$$\xi = \xi_0 [\ (T_f(\omega)-T_g)/T_g\ ]^{-\nu} \quad (3)$$

($\nu$ being the usual critical exponent for the correlation length in a phase transition). The dynamic scaling hypothesis states that the response time $\tau$ of a spin ensemble of size $\xi$ goes like

$$\tau \propto \xi^z \ , \quad (4)$$

where $z$ is defined as a dynamical exponent. In this way we obtain the critical dynamics scaling relation (Binder and Young, 1984) :

$$\tau = \tau_0 [\ (T_f(\omega)-T_g)/T_g\ ]^{-z\nu} \ . \quad (5)$$

The precise frequency dependence of $T_f(\omega)$ has been studied in many spin-glass examples. It clearly shows a divergence of the response times as $\omega$ goes to zero, attesting the critical character of the freezing process at $T_g$. The value of the exponent $z\nu$ ranges from 5 to 11 in the various samples (Bontemps et al, 1984; Vincent and Hammann, 1986; Kawamura and Taniguchi, 2015). This collective behaviour of the interacting spins at the spin-glass transition marks an important difference with other classes of freezing processes. As an example, for non-interacting superparamagnetic nanoparticles, the freezing process involves response times $\tau$ which follow an Arrhenius law

$$\tau = \tau_0 \exp(U/k_B T) \ . \quad (6)$$

The superparamagnetic fluctuations of the nanoparticles are freezing because of thermal slowing down due to their individual anisotropy barriers $U$. On the contrary, if the nanoparticles are concentrated enough to have significant magnetic interactions, they form a *superspin* glass that has also been well characterized (Dormann et al, 1996; Nakamae et al, 2012).

## 2.3 A genuine thermodynamic phase transition

In order to complete the characterization of a critical behaviour in the dynamics, it is interesting to check whether the spin-glass transition also presents a critical character for static quantities. In a ferromagnet, the order parameter is the spontaneous magnetization, and the order parameter susceptibility is the usual magnetic susceptibility. In a spin glass, defining an order parameter of the "glassy order" is not straightforward. A theoretical definition has been proposed by Edwards and Anderson (1975), in which the order parameter is the average over the sample of the squared moduli of the spins. The corresponding order parameter susceptibility has been shown to be the non-linear part of the magnetic susceptibility (Suzuki, 1977; Chalupa, 1977; Suzuki and Miyashita, 1981). The *dc* magnetic susceptibility $\chi$ can be expanded in even powers of the magnetic field $H$ :

$$\chi = \chi_0 - a_3 H^2 + a_5 H^4 \ldots , \qquad (7)$$

$\chi_0$ being the linear susceptibility. The coefficients of the non-linear terms are expected to diverge at $T_g$, with the critical exponents $\beta, \gamma$ corresponding to the spin glass order parameter :

$$a_3 \propto (T-T_g)^{-\gamma}, \ a_5 \propto (T-T_g)^{-(\beta+\gamma)}, \text{ etc.} \qquad (8)$$

Their determination implies extensive measurements of the magnetic susceptibility as a function of the field, at various temperatures close to $T_g$, and careful extrapolation to zero field. Figure 2 shows the example of Ag:Mn$_{0.5\%}$ (Bouchiat, 1986). The non-linear part of the susceptibility is plotted as a function of the field. As the temperature $T$ decreases towards $T_g$, a significant increase of the initial slope of the curves is clearly visible. Below $1.1 T_g$ = 2.97K, the low-field behaviour of the non-linear susceptibility goes from quadratic to singular. This increase (together with the analysis of other data) allows the characterization of the diverging character of $a_3, a_5$ (Eq. 7-8), and a coherent determination of the critical exponents $\beta, \gamma, \delta$. Similar results have been obtained in numerous different spin glass samples (see references in Kawamura and Taniguchi, 2015). Thus, the transition to the spin-glass phase has the same characteristics as a genuine thermodynamic transition to an ordered state.

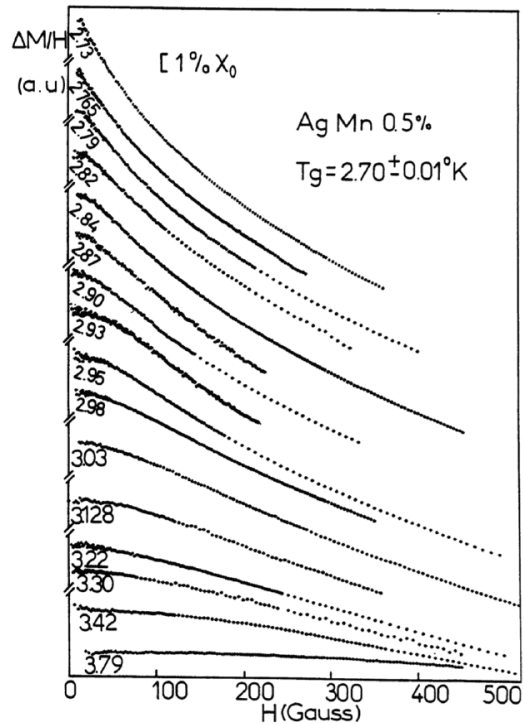

**Figure 2 :** From Bouchiat, 1986. Non-linear susceptibility (obtained as the difference $\Delta M$ between the total magnetization and its field-linear part, divided by the applied magnetic field $H$), as a function of field $H$, at different temperatures approaching the critical temperature $T_g$ =2.70K. The relative origins on the Y-axis are arbitrary. 1% of the linear susceptibility $\chi_0$ is indicated. As $T \to T_g$, a sharp increase of the slope at the origin is clearly visible, pointing to a divergence of the coefficients in the non-linear susceptibility (Eq.7-8).

The non-linear susceptibility can also be accessed through the study of the harmonics in the response to an *ac* field. Pioneering measurements of the 3$^{rd}$ harmonics have been performed by Miyako et al as early as in 1979. Lévy (1988) determined the whole set of critical exponents of AgMn samples by carefully measuring the 3$^{rd}$, 5$^{th}$ and even 7$^{th}$ harmonics of the *ac* susceptibility.

## 2.4 The spin-glass transition : open questions

The critical character of the spin-glass transition, which takes place in a fully disordered system, raises a number of issues. From the critical dimensions obtained in the mean-field theory of spin glasses, in $d$=3 a phase transition is expected for Ising (scalar) spins, but not for Heisenberg (vector) spins (Mezard, 1987, and references therein). However, a phase transition is found in real $d$=3 spin glasses, and it has been observed as well with Ising as with Heisenberg-like spins. A way to an explanation can be found in the scenario proposed by Kawamura of a chirality driven phase transition of spins (1-step-like RSB class model). This mechanism has been detailed and argued both analytically and numerically. A remarkable agreement of the predicted critical exponents with the experiments has been found (Kawamura, 1992; Kawamura and Taniguchi, 2015). A direct experimental access to chirality freezing is difficult, but some pioneering measurements of the anomalous Hall effect in spin glasses have already given very interesting results (Pureur et al, 2004; Taniguchi, 2007).

In the Parisi solution of the mean-field Ising spin glass (Parisi, 1979), an infinite number of different pure states is obtained, yielding a complex structure of the spin glass phase (this class of models corresponds to a continuous RSB, also called full-RSB, see in Mezard 1987, Altieri and Baity-Jesi 2023). It suggests that, after cooling from the paramagnetic phase, many domains with different types of spin-glass order should coexist and compete. A phase diagram with a transition line is obtained as a function of the magnetic field. On the other hand, with a very different point of view, scaling theories of the spin-glass behaviour have been developed for Ising spins (droplet model by Fisher and Huse, 1988a, 1988b; domain model by Koper and Hilhorst, 1988). In these models, there are simply two (spin reversal symmetric) pure states, and the phase transition is expected to be destroyed by any magnetic field. Let us briefly comment on these questions at the light of the experimental results.

(i) The question of a multiple or single nature of the ground state in the spin glass is very difficult to address directly in experiments, in which no ground state is ever reached. Some arguments are given in the discussion on the temperature dependence of aging effects (Sect. 4.3 below). A more direct way can be searched from the theoretical suggestion by Carpentier and Orignac (2008), stating that the correlation of the conductance fluctuations in two spin states should be a direct function of their overlap. A new experimental approach using transport measurements on mesoscopic samples has started (Capron et al, 2013). Measurements of the universal conductance fluctuations in mesoscopic spin glasses are challenging, but they can be a promising way to obtain information on the nature of the pure states. Resistivity measurements on mesoscopic wires are now on the way (Forestier et al, 2020).

(ii) The vanishing of the phase transition in presence of a magnetic field has been reported in a study of critical dynamics on a $Fe_{0.5}Mn_{0.5}TiO_3$ single crystal, which is a good example of a short-range Ising spin glass (Mattsson, 1995). Interestingly, recent experiments on $Dy_xY_{1-x}Ru_2Si_2$ show that the phase transition persists in a field in this example of an Ising but long-range (RKKY) system (Tabata et al, 2017). For Heisenberg-like spin glasses, data from torque measurements bring robust evidence for a true spin glass ordered state that survives under high applied magnetic fields (Petit et al, 2002).

The nature of the glass transition is also a hot topic for structural glasses. We now understand that the non-linear dielectric susceptibility plays a similar role as in spin glasses (Bouchaud and Biroli, 2005). It has been recognized for a long time that the drastic dynamical slowing down at the glass transition of the most "fragile" glasses can be compared with the critical slowing down observed in spin glasses (Souletie and Bertrand, 1991). Non-linear dielectric susceptibility measurements in glycerol and other viscous liquids have now allowed determining the growth of correlations at the approach of the transition and in the glassy phase during aging (Brun et al, 2012; Albert et al, 2019).

In experiments on spin glasses, what we see is essentially an out-of-equilibrium behaviour, with many interesting properties. We now describe some of them.

# 3. Slow dynamics and aging in spin glasses

## 3.1 Relaxation of the magnetization

After cooling the spin glass from above to below $T_g$ in a field, we reach a state in which the magnetization is approximately constant (FC state, Fig.1a). There are some signs that this frozen state is not at equilibrium. If we now turn the magnetic field to zero, the magnetization does not go to zero value, which for symmetry reasons should be the equilibrium value. The non-zero remanent magnetization obtained in this procedure, called *thermo-remanent magnetization* (TRM), slowly relaxes as a function of time. Such TRM curves are shown in Figure 3 on a logarithm time scale.

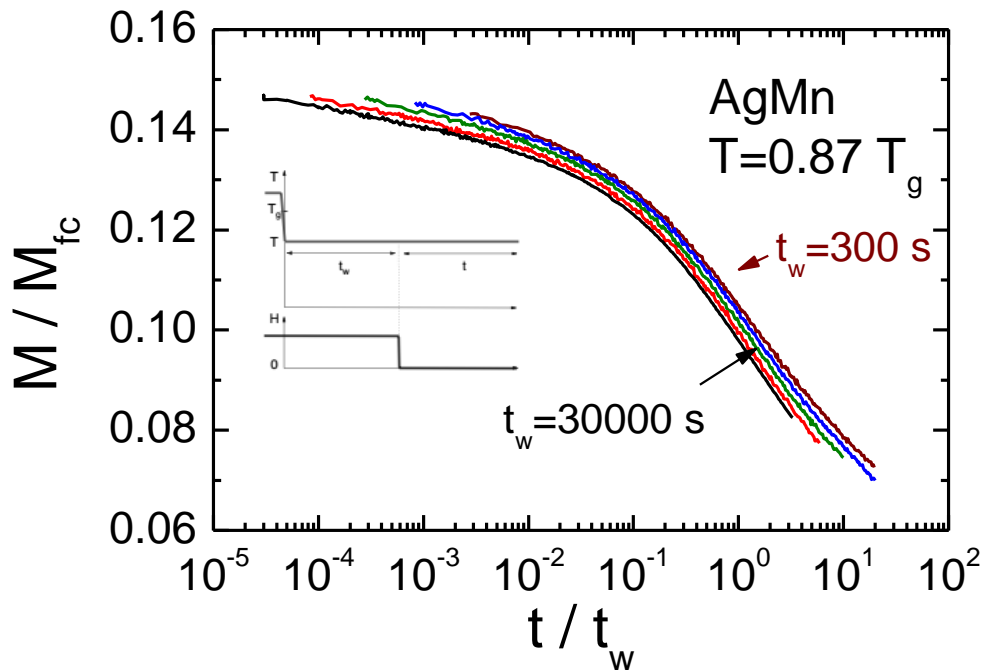

**Figure 3** (data from the Saclay group, figure from Vincent et al, 1997). Relaxation of the thermo-remanent magnetization (TRM) of the Ag:Mn$_{2.7\%}$ spin glass. The procedure is sketched in the inset. The magnetization decay curves obtained for 5 $t_w$ values ranging from 300 s to 30000 s are presented on a logarithmic scale, as a function of $t/t_w$ to emphasize their $t_w$-dependence (called "aging").

These experiments involve a second time parameter called the *waiting time* $t_w$ (inset of Fig.3). The sample is rapidly cooled in a weak field *H= 0.1 Oe* from above $T_g$ to $T=0.87T_g$ (quench), and is kept at temperature *T* during a given $t_w$ (5 values from 300 to 30 000 s in Fig.6). After waiting $t_w$, the field is cut, and the relaxation is measured as a function of the observation time *t* elapsed since the field cut-off. The inflection point of curves (maximum of the relaxation rate) shifts towards longer times as $t_w$ increases, always taking place around $t \approx t_w$. Presented as a function of the ratio $t/t_w$, they are almost superimposed. So, after a longer waiting time $t_w$, the response to the field change becomes slower, with characteristic times of the order of $t_w$ itself. The spin-glass phase has been evolving during the time spent below $T_g$ (it has become "more frozen"), although the FC-magnetization remains roughly constant. This phenomenon is called *aging* (from similar effects observed in polymer mechanics, Struik, 1978).

When studied in more details, the $t_w$ effect can be accurately taken into account. The shift between the curves obtained for different $t_w$'s is not exactly *log $t_w$*, but rather $\mu.\log t_w$, $\mu$ being a sample-dependent coefficient of order 0.8-0.9 . Hence, the relevant variable is $t_w^\mu$, and when it is properly accounted the relaxation curves can all be superimposed very precisely (Ocio et al, 1985a; Vincent et al, 1997; see an example in the inset of Fig.5 below).

We want to emphasize that the magnetic field applied in these experiments is weak (0.1-10 Oe), that is, not larger than a few percent of the value that, at low temperature, would bring the spin glass back to the paramagnetic phase. Such a weak field has a negligible effect on the properties that we describe here, linear response is obeyed, the field can be considered as a non-disturbing probe field. This will not hold for higher values of the field (see Fig. 2.8 in Vincent, 2007). Interestingly, some differences in the dynamics of the spin-glass phase with and without a field have recently been characterized very close to $T_g$, both in experiments and numerical simulations (Paga et al, 2022). In the field and temperature range of the measurements presented here, no difference was visible within the experimental accuracy.

It can also be useful to clarify that aging during $t_w$ is not related to the presence of the field. The mirror procedure of the TRM experiment can be performed with the same conclusions. Cooling the sample in zero field from above $T_g$, waiting $t_w$ at low temperature, and applying a field in the ZFC state yields a *ZFC(t)* relaxation which is the mirror image of the *TRM(t)* curve, and presents the same $t_w$-dependence. It has been demonstrated that *ZFC($t_w$,t) + TRM($t_w$,t) = FC* (this relation even holds when a slight relaxation of the FC magnetization can be seen, *FC ≡ FC(t)*) (Djurberg et al, 1995).

The phenomenon of aging has been known for a long time in the mechanical properties of a wide class of materials called "glassy polymers". When a piece of e.g. PVC is submitted to a mechanical stress, its response (elongation, torsion ...) is logarithmically slow. And the response depends on the time elapsed since the polymer has been quenched below its freezing temperature. Like in spin glasses, for increasing aging time the response becomes slower. The $t_w^\mu$-dependence of the dynamics of glassy polymers has been expressed as a scaling law (Struik, 1978) that could be applied successfully to the case of spin glasses (Ocio et al, 1985a). Numerous glassy materials show similar aging phenomena. Numerical simulations of packed hard spheres provide us with very powerful toy models of simple glasses (Jin and Yoshino, 2017).

## 3.2 Magnetic noise

We have shown above how aging in spin glasses can be studied by measuring the waiting time dependence of the magnetic response to cutting a magnetic field at low temperature (TRM relaxation). In statistical mechanics, the response to a field change is related to the spontaneous fluctuations by the Fluctuation-Dissipation theorem (FDT), established for ergodic systems at equilibrium (see references in Hérisson and Ocio, 2002, 2004). Applying FDT to the situation of TRM experiments, we can relate the relaxation function $\sigma(t',t)$ ($\sigma = M/H$, magnetic response at *t* after cutting off a field *H* at *t'*) to the autocorrelation *C* of the fluctuations of the magnetization *m*, namely *C(t',t)=<M(t').M(t)>* :

$$\sigma = C/k_BT , \quad (9)$$

with $k_B$ being the Boltzmann constant.

The measurement of the spontaneous fluctuations of the magnetization ("magnetic noise") in spin glasses is very difficult because of a very weak signal. Yet, it has been a strong motivation for the experimentalists to check how far FDT can be obeyed in the out-of-equilibrium situation of the spin glass, for which the aging regime is considered archetypal. Pioneering measurements started early (Ocio et al, 1985b). Later on, important theoretical efforts have been devoted to extensions of FDT to non-equilibrium situations (see for example Franz et al., 1998). A prominent result by Cugliandolo and Kurchan (1994,1997) is a modified FDT relation which reads

$$\sigma = C.F(C)/k_B T \quad (10)$$

where *T/F(C)* takes the meaning of an *effective temperature* $T_{eff}$, different from the sample temperature *T*. In this approach, for large *t'*, the obtained correction factor *F(C)* is a function of the autocorrelation *C* only, i.e. it does not explicitly depend on *t* and *t'*, but has a time dependence through the value of *C(t',t)* only. Hérisson and Ocio (2002, 2004) have built a special purpose SQUID magnetometer in which response and fluctuations could both be measured in the same geometry. Important care was taken for eliminating all electromagnetic parasite sources.

An example of noise recordings is presented in Figure 4, from the $CdCr_{1.7}In_{0.3}S_4$ thiospinel sample. Each trace shows the SQUID output (proportional to the sample magnetization, with an arbitrary offset) during one experiment, starting from above $T_g$ and cooling. Due to a slight residual field, the trace shows the magnetization peak observed when crossing $T_g$. After cooling, the temperature is stabilized at *T=0.7$T_g$*, and the magnetization fluctuations are recorded as a function of time during ~$10^4$ s. After that the sample is re-heated above $T_g$. The experiment was repeated ~300 times. On each of the recorded traces, for any choice of times *($t_w$,t)* the correlation *M($t_w$).M($t_w$+t)* can be computed. This value is strongly fluctuating from one experiment to the other, but the average over ~300 measurements is taken and after properly subtracting offsets the noise autocorrelation *C($t_w$,t)=<M($t_w$).M($t_w$+t)>* is obtained.

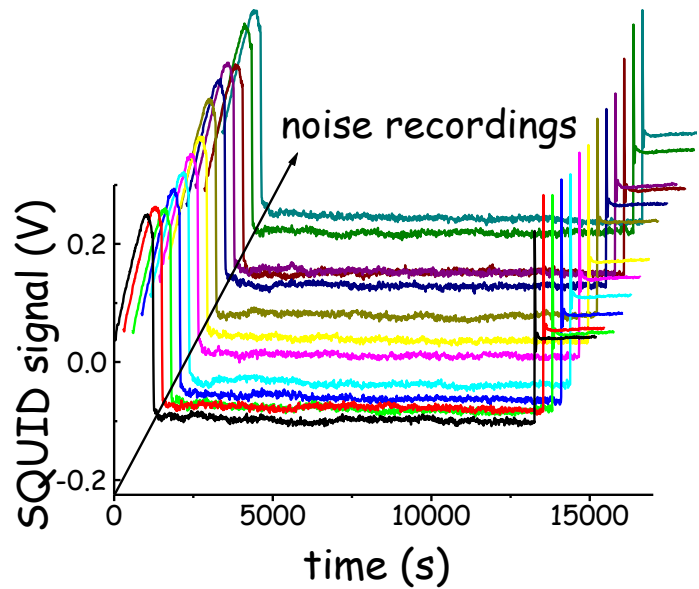

**Figure 4** (data from Hérisson and Ocio, 2004) The signal from the SQUID detector, proportional to the magnetization with an arbitrary offset, is shown as a function of time in a series of successive noise recording experiments. The spontaneous fluctuations of the magnetization (magnetic noise) are observed here in the absence of any excitation.

Figure 5 (bottom part) shows TRM experiments performed in the same setup, with a very small excitation field of ~$10^{-3}$ Oe. Figure 5 (top part) shows the noise autocorrelation, presented in the same way as the TRM results, as a function of $t$ for various fixed values of $t_w$.

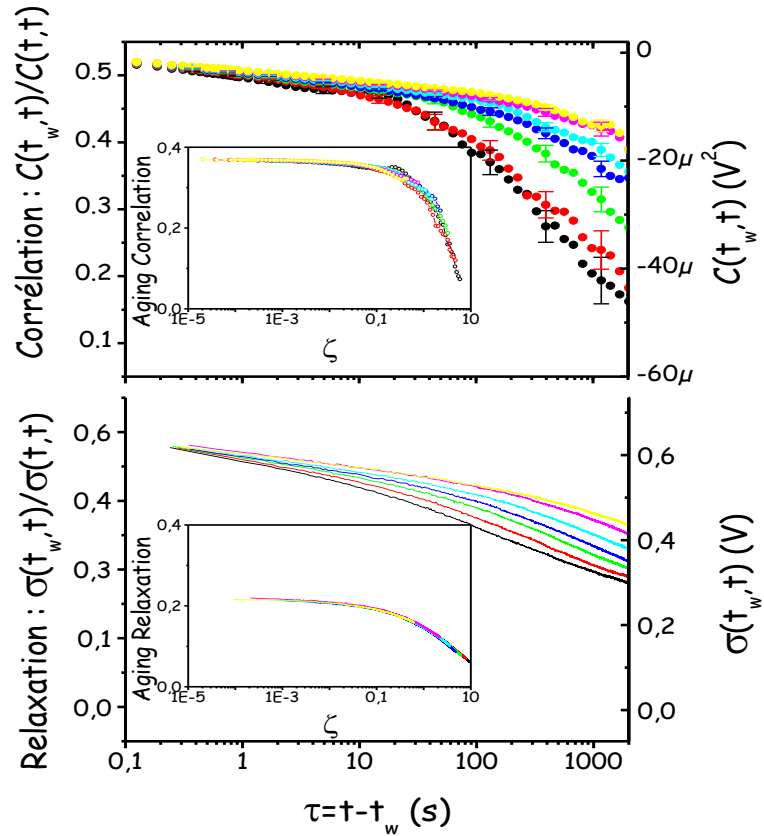

**Figure 5** (data from Hérisson and Ocio, 2004) TRM-relaxation (**bottom**), and noise autocorrelations functions presented in the same fashion as the TRM curves (**top**). The different curves correspond, in ascending order, to 5 values of the waiting time $t_w$, increasing from 100 to 10000 s. The insets show a superposition of the aging part of the curves using the scaling procedure of Ocio et al, 1985a (see also Vincent et al, 1997). This double figure allows an overall comparison of the response and autocorrelation functions, which are related by the Fluctuation-Dissipation Theorem (FDT). The experiment aims at testing FDT in the non-equilibrium conditions of aging.

Both sets of results present the same overall shape, but a precise comparison is best illustrated in the plot of Figure 6 (sometimes called "FDT plot"), in which the response function $\sigma(t_w,t)$ (or the susceptibility $\chi=1-\sigma$) is plotted as a function of $C(t_w,t)$ for 3 different temperatures $T=0.6, 0.8$ and $0.9T_g$. For each of the 3 temperatures, the point cloud is the set of "raw" results obtained for various values of $(t_w,t)$. The straight lines with $1/T$ slope show the prediction for classical FDT with no correction. There is a clear $1/T$ regime for the higher values of $C$, and a crossover towards a weaker slope $1/T_{eff}$ with $T_{eff}>T$ as $C$ decreases. These deviations show the first experimental observation in a spin glass of deviations from the normal FDT in the aging regime (Hérisson and Ocio, 2004).

The results in Figure 6 strongly suggest that the correction factor *F(C)* to FDT is only a function of *C*, as predicted by Cugliandolo et al (1994,1997). They allow a comparison with different models of the spin-glass phase, provided that a long-time extrapolation of the data can be made (see more details in Hérisson Ocio, 2004). The observed mean slopes correspond to $T_{eff}(0.6T_g)\sim1.5T_g$, $T_{eff}(0.8T_g)\sim3T_g$, and $T_{eff}(0.9T_g)\sim4T_g$. An infinite $T_{eff}$ value (horizontal slope of the data), as expected in domain growth type models (Barrat, 1998) like the droplet model (Fisher and Huse, 1988a, 1988b), seems to be very unlikely. In models like the mean-field spin glass with continuous RSB (Mezard et al 1987), a $\chi=(1-C)^{1/2}$ behaviour is predicted. The dashed line in Figure 6 shows a $\chi=(1-C)^{0.47}$ fit (Marinari et al, 1998) which gives at least a rough account of the results. However, in their 2004 paper, Hérisson and Ocio comment on the possibility of also analysing the results on the basis of a 1-step RSB model, confirming a possible relevance of the chiral scenario proposed by Kawamura (1992).

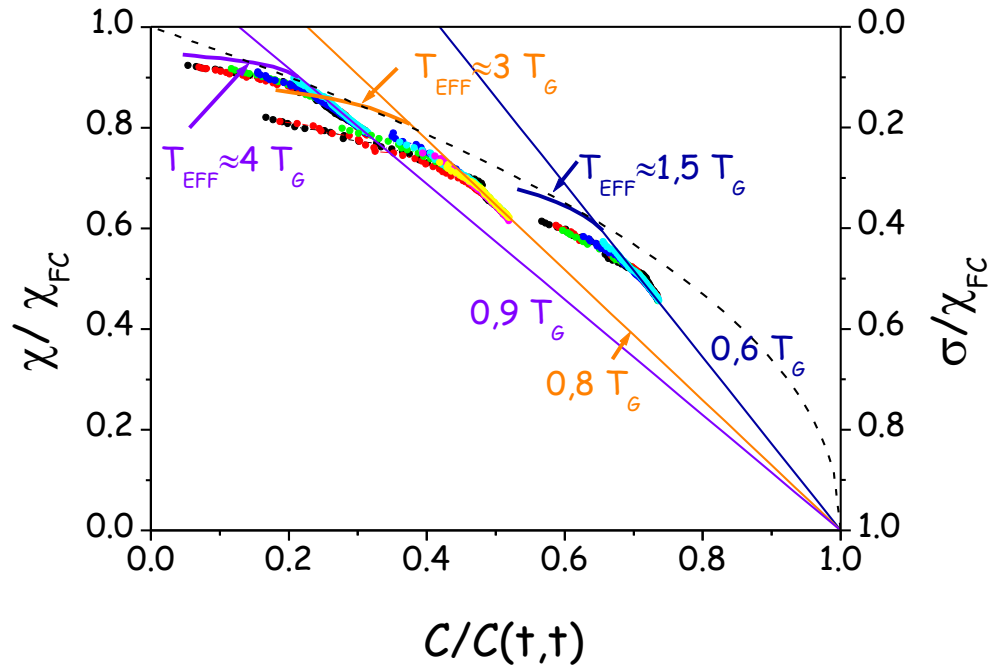

**Figure 6** (from Hérison and Ocio, 2004) Response function versus autocorrelation (so-called "FDT plot"), for 3 different temperatures *T*. The straight lines show the *1/T* equilibrium regime of FDT. The points are the raw results. The solid curves are a long time extrapolation of the data, which is necessary for a comparison with the mean-field theory. The dashed line is a fit to $\chi=(1-C)^{0.47}$, in reference to the continuous RSB model (Marinari et al, 1998).

## 4. Aging, Rejuvenation and Memory Effects

### 4.1 Temperature step experiments

The cooling rate has a strong influence on the state of a structural glass : slower cooling yields smaller values of the enthalpy and specific volume, which are closer to equilibrium values (Struik, 1978; Simon and Mac Kenna, 1997; Ediger and Harrowell, 2012). In the case of spin glasses, can we play with the cooling rate, or design clever temperature procedures, in order to bring the spin glass closer to equilibrium ? This question was the starting point of a class of experiments (Refregier et al, 1987), which have led to the observation of the astonishing phenomena of rejuvenation and memory (see a review in Vincent, 2007; Vincent and Dupuis, 2018).

Aging can also be observed in *ac* susceptibility measurements (like in Fig.1b). After cooling from the paramagnetic phase to some fixed temperature in the spin-glass phase, both components $\chi'$ and $\chi''$ are seen to slowly relax downwards due to aging. The variation in relative value is more important for the out-of-phase susceptibility $\chi''$ (Dupuis, 2002; Vincent and Dupuis, 2018), therefore most experiments have focused on the measurement of $\chi''$. Figure 7 presents an *ac* experiment in which temperature steps have been applied during aging (Lefloch et al, 1992).

After a normal cooling (∼100 s) from 1.3 $T_g$ to 0.7 $T_g$ = 12K ($T_g$ = 16.7K), the sample is kept at the constant temperature of 12K for $t_1$ = 300 min. During this time, $\chi''$ shows a strong downwards relaxation (aging). After 300 min, the temperature is lowered one step further from $T$ = 12 K to $T$-$\Delta T$ = 10 K. What is then observed is a $\chi''$ jump and a renewed relaxation. This behaviour upon further cooling was termed *rejuvenation*, because the restart of the relaxation suggests that aging is starting anew at $T$-$\Delta T$. Apparently, there is no influence of former aging at $T$. The rejuvenation effect is at variance with the common expectation that a slower cooling rate should help approaching equilibrium. Waiting $t_1$ at $T$ did not help approaching equilibrium at $T$-$\Delta T$.

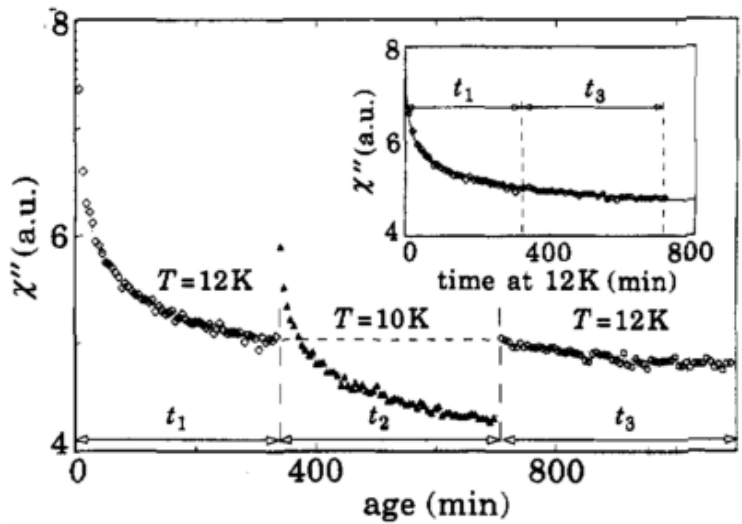

**Figure 7** (from Lefloch et al, 1992) AC experiment at 0.01 Hz, relaxation of the out-of-phase susceptibility $\chi''$ during a negative temperature cycle following a temperature quench. The points show aging at 12K, rejuvenation at 10 K, and memory at 12 K. The inset shows that, despite rejuvenation at 10 K, both relaxations at 12K are in continuation of each other (memory). The sample is the CdCr$_2$xIn$_{2(1-x)}$S$_4$ thiospinel spin glass ($T_g$=16.7K).

One may naturally ask whether this renewed relaxation corresponds to a full rejuvenation of the sample. The answer is no. Two remarks should be made at this point :

(i) Clearly, for continuity reasons, the amount of rejuvenation must be modulated by the value of $\Delta T$. A restart cannot be expected after a tiny temperature variation. The temperature interval must be sufficiently large, here we need ΔT ≥ 2 K.

(ii) We should keep in mind that the time window explored in an experiment is limited, whereas the aging state of the spin glass involves relaxation processes on a very wide time range. Hence, from this sole experiment we do not have the knowledge of the overall state at all time scales.

The final part of the measurement completes the answer to the question of the occurrence of a full rejuvenation. After observing the renewed relaxation during $t_2$ = 300 min at $T$-$\Delta T$ = 10 K, the temperature is turned back to $T$ = 12 K. Then, what we see is that $\chi''$ goes up to the point that was attained at the end of the stay at the original temperature $T$. And the relaxation continues in precise continuity of the former one, as if nothing of relevance for the state at $T$ had happened at $T$-$\Delta T$. This can be checked by shifting the third relaxation to the end of the first one (inset of Fig. 7): they are in continuity, and can be superposed on the reference curve which has been separately measured in simple aging at constant $T$ = 12K.

Hence, during aging at $T$-$\Delta T$ and despite the strong associated $\chi''$-relaxation, the spin glass has kept a *memory* of previous aging at $T$. This memory is retrieved when heating back to $T$. This experiment made with $\Delta T$ = 2 K illustrates in a spectacular manner the phenomenon of rejuvenation and memory in a spin glass. For smaller $\Delta T$ values, the situation is obviously more complex : one finds a weaker rejuvenation, and some mutual influence of aging at both temperatures ("cumulative" aging at both temperatures, similar to a cooling rate effect). Details on the results in this regime of small temperature variations, together with their discussion in terms of a random energy model, can be found in Sasaki et al, 2002. A wide set of experiments have been performed by the Saclay group (see references in Vincent, 2007) and by the Uppsala group (see, for example, Jönsson et al, 2002), with similar results, although sometimes discussed in slightly different terms regarding the comparison with the droplet model (Fisher and Huse, 1988a, 1988b).

Very recently, Zhai et al (2022) could accurately measure the effect of temperature variations on the tiny relaxation (≈0.5%) of the FC magnetization in a CuMn spin glass. They can precisely distinguish a regime of cumulative aging for small temperature variations, and another one obtained for larger temperature variations, which shows some rejuvenation of the relaxation and is discussed in terms of "temperature chaos", a theoretical notion which corresponds to an extreme sensitivity of the glass to temperature changes (Bray and Moore, 1987).

## 4.2 Multiple memories

The ability of the spin glass to keep a memory despite rejuvenation has been further explored in experiments with multiple temperature steps. The first *memory dip* experiments, suggested by P. Nordblad, were developed in a collaboration between the Uppsala and Saclay groups (Jonason et al, 1998; Jonason et al, 2000). An example of a "multiple dip" experiment is shown in Fig. 8 (Bouchaud et al, 2001; Dupuis, 2002). This is a $\chi''$ measurement in which the sample is cooled from above $T_g$ to 4K by steps of 2 K, with a stop at each step. Then, the sample is reheated continuously (inset of Fig. 8). Starting from $T>T_g$ where $\chi''$=0 (paramagnetic phase), $\chi''$ rises up when crossing $T_g$ = *16.7 K*, and when the cooling is stopped at 16K, 14 K, etc., the relaxation of $\chi''$ due to aging is recorded during 30 minutes.

After waiting 30 minutes at constant temperature, upon further cooling by another 2 K step, a $\chi''$ jump of rejuvenation is found, and the relaxation due to aging again takes place. At each new cooling step, rejuvenation and aging are seen, and this happens ~ 6 times in the experiment of Fig. 8.

In the second part of the experiment, the sample is re-heated continuously at a slow rate (0.001 *K/s*, equal to the average cooling rate). Amazingly, apart from the rather noisy low-*T* region, the memory of each of the aging stages performed during cooling is revealed in shape of "memory dips" in $\chi''(T)$, tracing back the lower value of $\chi''$ which was attained at each of the aging temperatures. Thus, the spin glass is able to keep the simultaneous memory of 5 or 6 aging stops in a row, performed at lower and lower temperatures. The curve obtained while increasing the temperature after that reveals the memories (and meanwhile erases them).

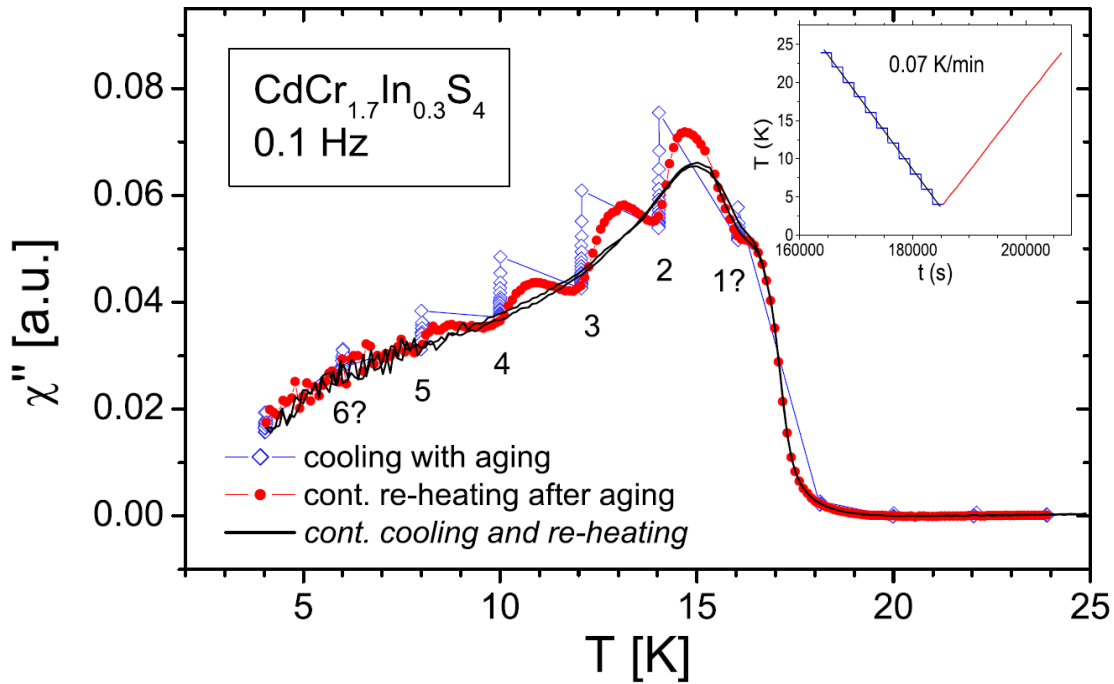

**Figure 8 :** (from Dupuis, 2002; Bouchaud et al, 2001) An example of multiple rejuvenation and memory steps. The inset is a sketch of the procedure. The sample was cooled by 2 K steps, with an aging of time of 30 minutes at each step, after which rejuvenation can be seen (open blue diamonds). Continuous reheating at 0.001 K/s (full red circles) shows memory dips at each temperature of aging. The solid black line shows a reference measurement with continuous cooling/heating and no steps.

## 4.3 Domain growth and hierarchy of states

In the droplet model (Fisher and Huse 1988a, 1988b), the spin glass is thought in the same terms as a ferromagnet, having two symmetric equilibrium states of opposed spin directions (with no trivial ordered alignment of the spins). Aging is described as the slow growth of domains of this "spin-glass order" of hidden symmetry, starting from the random configuration created by the quench from the paramagnetic phase. In a naive interpretation, domain growth processes should progress continuously during temperature variations, with some influence of temperature on the growth rate, similar to cooling rate effects. But, in that case, it is difficult to imagine how rejuvenation and memory effects may arise. In the droplet model, they are related to "temperature chaos" effects (Bray and Moore, 1987). Discussions on the relevance of this scenario to experiments can be found in Yoshino et al, 2001, and Jönsson et al, 2004. The notion of a growing size of cooperative regions, encountered in many glassy systems, as well in experiments as in theoretical models, may well be at the heart of aging phenomena, as detailed by Corberi et al (2011).

The theoretical ingredients that are necessary for rejuvenation and memory effects to arise are not yet well determined. Early studies of frustrated magnetic systems have shown that average effective interactions can be defined, which may present a strong temperature dependence (Miyashita, 1983). The temperature variation of the effective interactions can cause rejuvenation effects (Miyashita and Vincent, 2001). Still, the preservation of memory despite rejuvenation requires considering other processes in addition (Tanaka and Miyashita, 2005).

On the other hand, a different point of view has been proposed by the Saclay group to account for the rejuvenation and memory effects. The guideline is to consider a hierarchical organization of the metastable states as a function of temperature, inspired by the hierarchical organization of the pure states in the Parisi solution of the mean-field spin glass (Mezard et al, 1987). Indeed, it could be shown that rejuvenation and memory effects can be expected in the dynamics of this model (Cugliandolo and Kurchan, 1999).

In this scheme, pictured in Fig. 9 (Dotsenko et al, 1990; Refregier et al, 1987), the effect of temperature changes is represented by a modification of the free-energy landscape of the metastable states (not only a change in the transition rates between them). At fixed temperature $T$, aging corresponds to the slow exploration of the numerous metastable states (at level $T$ in Fig. 9). When going from $T$ to $T-\Delta T$, the free-energy valleys subdivide into smaller ones, separated by new barriers (level $T-\Delta T$ in Fig. 9). Rejuvenation arises from the transitions that are now needed to equilibrate the population rates of the new sub-valleys: this is a new aging stage. For large enough $\Delta T$ (and in the limited experimental time window), the transitions can only take place between the sub-valleys inside the main valleys, in such a way that the population rates of the main valleys are untouched, keeping the memory of previous aging at $T$. Hence the memory can be retrieved when re-heating and going back to the $T$-landscape.

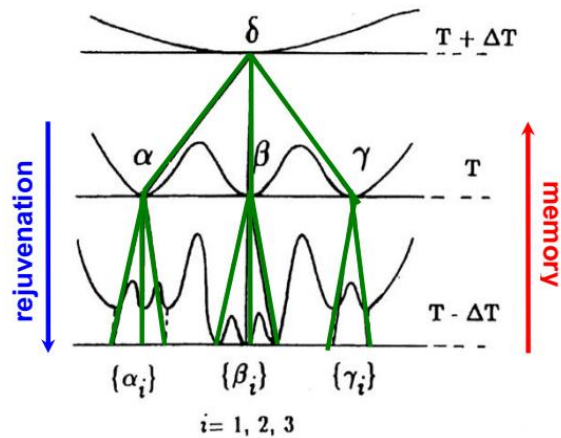

**Figure 9 :** Schematic picture of the hierarchical structure of the metastable states as a function of temperature (from Dotsenko et al, 1990; Refregier et al, 1987).

This tree picture allows a qualitative description of the rejuvenation and memory effects, but it is also able to reproduce quantitatively many features of the experiments (see discussions of the effect on aging of small temperature changes in Sasaki et al, 2002; Vincent, 2007). Tree models, starting with Derrida's generalized random energy model (GREM, Derrida and Gardner, 1986) and Bouchaud's trap model (Bouchaud and Dean, 1995), offer a fruitful framework for modelling out-of-equilibrium systems (Hoffmann and Sibani, 1989; Sasaki and Nemoto, 2001). Samarakoon et al (2017) have investigated memory effects in various magnetic glassy materials (high-temperature superconductors, spin-orbit Mott insulators, frustrated magnets, and a metallic spin glass). They observed memory effects, sharper in the spin glass, less pronounced and closer to a cooling rate effect in other systems. By comparing with a model of random walk on a tree structure, the authors show that the memory effect depends on the hierarchical character of the tree. Recently, Zhang et al (2021) characterized a disappearance of the glassy behaviour and the memory effect above a critical value of tree branching ratio.

The hierarchical scheme in Fig. 9 shows free-energy barriers that are growing as the temperature decreases. The slowing down observed in the dynamics is indeed more drastic than expected from thermally activated dynamics in a fixed landscape. In a series of experiments, the growth of the free-energy barriers upon lowering temperature has been measured (Hamman et al, 1992). The results suggest that some barriers are diverging at all temperatures below $T_g$, making a link between the hierarchical organization of the metastable states as a function of temperature and that of the pure states in the mean-field spin glass model (a slightly different barrier analysis is presented in Bouchaud et al, 2001).

Let us try to merge into a common framework both views of aging phenomena that we have considered in this discussion :
- a domain growth picture (suggested by the slowing down of the dynamical response during aging), and
- a hierarchical organization of some entities to be defined (suggested by rejuvenation and memory effects).

When large domains have been established during aging at temperature $T$, with size $L_T$, rejuvenation at $T$-$\Delta T$ implies that new processes take place at $L_{T-\Delta T} < L_T$. In order to ensure the preservation of memory at $T$, we need $L_{T-\Delta T} \ll L_T$, together with a freezing of the $L_T$ processes at $T$-$\Delta T$. This is the "temperature-microscope effect" proposed by J.-P. Bouchaud (Bouchaud et al, 2001). In experiments like those from Figures 7,8 (and 10), at each temperature stop aging should take place at well-separated length scales $L_n < ... < L_2 < L_1$, as if the magnification of the microscope were strongly varied at each temperature step.

This "hierarchy of embedded length scales" as a function of temperature is a real space equivalent of the hierarchy of metastable states in the configuration space (Fig. 9). In such a scheme, the domains are not compact, they are embedded cooperative entities in which the spins are correlated up to a characteristic length, which varies quickly with temperature (the crucial point for memory is that the characteristic times increase quickly with decreasing temperature). Later below, in Section 5, we describe experiments in which the correlation length of the spin-glass order can be measured, which bring more information about these embedded length scales.

## 4.4 An experiment gathering the various aspects of aging

We now show an experiment in which cooling rate effects, illustrative of domain-growth processes, can be observed together with memory phenomena.

This is a ZFC-type procedure (*dc*, like in Fig.1a), proposed by the Uppsala group (Mathieu et al, 2001). The sample is cooled in zero-field with several different thermal procedures, and after applying the field at low temperature the ZFC magnetization is measured while increasing the temperature continuously at fixed speed (small steps of 0.1K/min).

The results are shown in Figure 10. Firstly, we can observe the effect of a slow cooling in comparison with that of a fast cooling : the ZFC-curve obtained after a slow cooling lies below the one obtained after a fast cooling. That is, after a slower cooling, the response to the application of the field at low temperature is weaker, the spin-glass state has been more robustly established : the growth of spin-glass ordered domains has been favoured by the slower cooling.

Secondly, memory effects can also be observed in this experiment, using now a procedure in which the cooling is interrupted at several temperatures during a one hour waiting time (like in the $\chi''$ experiment in Fig. 8). The magnetization measured during re-heating after this step-cooling procedure shows clear dips at all temperatures at which the sample has been aging. These effects are emphasized in the right part of Figure 10 where we have plotted the difference between the curves obtained after a specific cooling history and the reference one obtained after a fast cooling. Sharp oscillations (memory dips) show up on top of a wide bump (cooling rate effect).

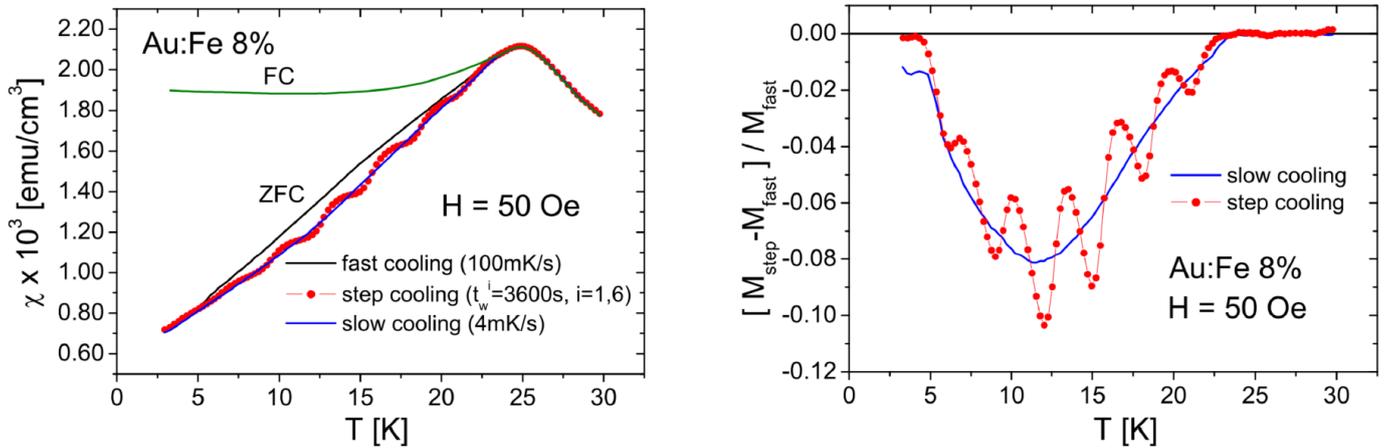

**Figure 10 :** (from Dupuis, 2002) **Left :** Effect of various cooling procedures on the ZFC susceptibility (magnetization divided by field) of the Au:Fe8% spin-glass. Comparison of the effect of fast and slow cooling, with and without stops. **Right :** difference with the magnetization obtained after fast cooling, emphasizing the memory dips on top of a cooling rate effect.

We may wonder whether the structural glasses, usually considered to be dominated by cooling rate effects, also present rejuvenation and memory phenomena. Experiments have been designed to search for such effects in various glassy systems (including polymers, gels…). They have indeed identified comparable phenomena, although usually less marked than in spin glasses (e.g. Bellon et al, 2002; see references in Vincent and Dupuis, 2018). In simulations of polydisperse repulsive particles, Scalliet and Berthier (2019) have shown that rejuvenation and memory effects are found for packing fractions corresponding to soft colloids and granular materials, whereas such effects are not seen in the regime describing dense supercooled liquids like structural glasses.

An interesting example concerns the mechanical properties of gelatine (Parker and Normand, 2010). Gelatine is a complex protein made of folded helices, and it has many degrees of freedom related to helix unfolding in the vicinity of room temperature. The experiment that we show in Figure 11 is an *ac* measurement of the elastic modulus $G'$, it can be compared with the *ac* experiment on a spin glass in Fig. 8. $G'$ has first been measured during simple cooling and re-heating (dashed line, reference). Another measurement has been performed with two stops at two different temperatures during cooling (solid line). During each stop, $G'$ relaxes upwards (the gelatine stiffens, this is aging), and upon further cooling some rejuvenation can be seen. When re-heating, $G'$ shows a dip at each of both stop temperature, the two memory dips are overlapping, but they can be clearly distinguished.

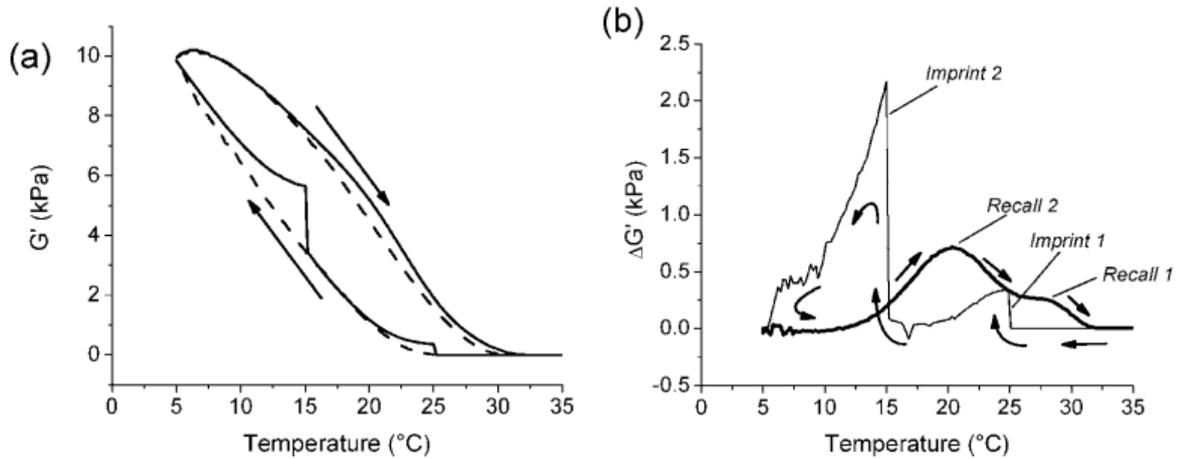

**Figure 11** (from Parker and Normand, 2010) Measurement of the elastic modulus *G'* of a gelatine gel, showing aging and memory effects during temperature cycles. Two stops of 2 hours were made at 25 and 15°C during cooling. The left part **(a)** shows *G'* as a function of temperature, in solid line for the cycle with stops, and in dashed line for a cycle without stops. During the stops, *G'* increases slowly (aging). Upon re-heating, a widespread excess of *G'* is seen when compared with the curve obtained without stops. In the right part **(b)** of the figure, which shows the difference plot, the memory of both stops shows up clearly.

# 5. A correlation length for spin-glass order

## 5.1 The spin-glass order made measurable

It is not obvious to observe the spin-glass ordered domains, since these correlated entities do not present any macroscopic symmetry. Still, the existence of correlated domains separated by frustrated regions has already been mentioned ("clusters") in early studies of the ±*J* spin-glass model (Miyashita and Suzuki, 1981), and small scale simulations allowed the visualization of some long-living domains by studying the spin autocorrelation function (Takano and Miyashita, 1995). Also, in their simulations of the Heisenberg spin glass, Berthier and Young (2004) studied the relative orientations of the spins in two copies ("replicas") of the system which, starting from different random states, evolve independently by a Monte-Carlo algorithm. Regions of the sample where the spins have a constant angle between both replicas can be rather clearly distinguished. These replica-correlated regions are in a sense equivalent to spin-glass ordered domains, and their growth as a function of time is visible (Fig. 5 in Berthier and Young, 2004).

Yet, it is a challenge to observe these domains in real spin-glass materials. Starting from an idea of R. Orbach, experiments could be developed which allow the measurement of the correlation length of growing domains during aging (Joh et al, 1999). Until this point, we presented experiments which were performed at very low magnetic field, i.e. in which the field is probing but not influencing the spin-glass state (as explained in Sect. 3). When using higher amplitudes of the field, we observed that the magnetization relaxation following a field change becomes faster (this can be seen as well in TRM as in ZFC procedures).

The TRM-relaxation curves can be characterized by their inflection point, $t_i$, which at low field takes place at $t_i \approx t_w$ (TRM curves in Fig. 3). According to thermally activated dynamics, the time $t_i$ defines a typical free-energy barrier $U$ that can be overcome at temperature $T$ after a time $t_i$ (with an attempt time $\tau_0 \approx 10^{-12}$s defined by the paramagnetic fluctuations):

$$U(t_i) = k_B T \, Ln(t_i/\tau_0) \qquad . \qquad (11)$$

For a low field TRM relaxation as in Fig. 3, $t_i \approx t_w$, but for higher field amplitudes the relaxation is accelerated and the inflection point $t_i$ shifts towards shorter times (Joh et al, 1999, 2000). Hence, as the field increases, the characteristic barrier $U(t_i)$ decreases ($t_i$ becomes smaller than $t_w$). We ascribe this energy decrease $U(t_w)-U(t_i)$ to the increase with the field of the Zeeman energy $E_Z$ which couples the magnetic field to the spins. In terms of spin-glass ordered domains involving a typical number of correlated spins $N_s$ with total magnetization M, $E_Z$ = M.$\mu_0$.H (here, M is an extensive quantity, not per unit volume as is $M$ in the rest of the paper; $\mu_0$ is the magnetic permeability of vacuum). A natural hypothesis is M $\propto N_s$, but for small numbers M $\propto N_s^{1/2}$ (typical fluctuation for $N_s$ randomly aligned spins) has also been considered (Bert et al, 2004). The recent comparisons of experiments and simulations now offer a better understanding of this question (Zhai et al, 2020; Paga et al, 2021). Let us simply consider here M $\propto N_s$.

For TRM experiments at a given $t_w$, measuring the field dependence of the inflection point of the curves allows the determination of $N_s(t_w)$. Repeating the measurements for various $t_w$ values yields the time variation of $N_s$. Figure 12 shows results obtained from different spin-glass samples (Joh et al 1999, 2000; Bert et al, 2004).

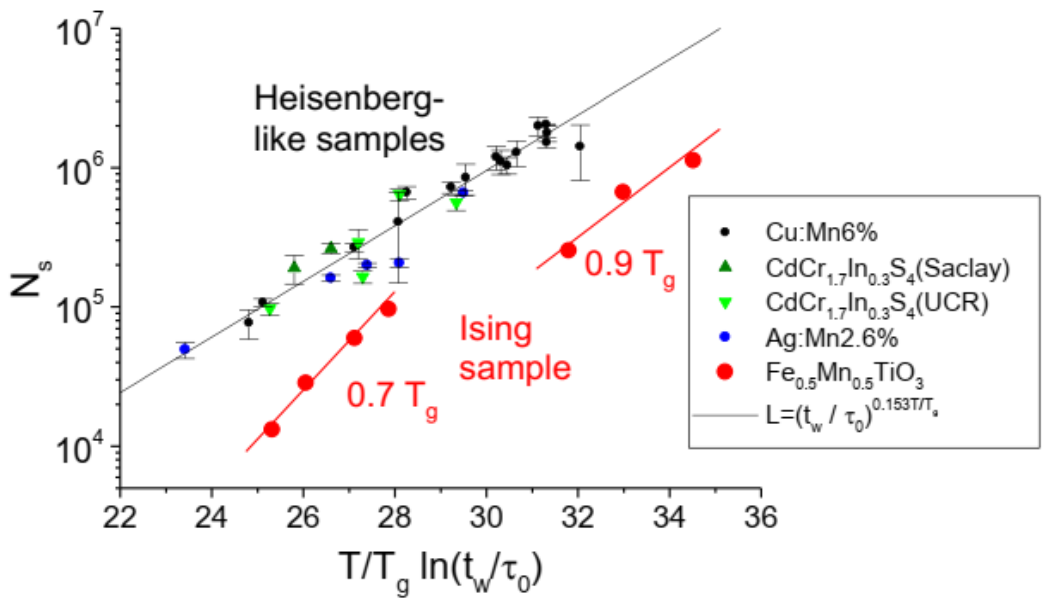

**Figure 12** (from Joh et al, 1999, 2000; Bert et al, 2004) Number of correlated spins extracted from field change experiments, as a function of the reduced variable $T/T_g \ln(t_w/\tau_0)$. The points with error bars correspond to Heisenberg-like spin-glasses, they are well fitted by the straight line $N_S \sim (t_w/\tau_0)^{0.45T/T_g}$. The full circles lying below the others are from the $Fe_{0.5}Mn_{0.5}TiO_3$ Ising sample and call for a more detailed discussion (Bert at al, 2004; Paga et al, 2021).

In this plot, the results from 3 samples do all fall on the same line. The number of correlated spins is, as expected, an increasing function of $t_w$, reaching during the experiments $10^4$-$10^6$, which means a range of 10-100 lattice units for the correlation length (considering crudely that the size $L$ goes like $N^{1/3}$). The 3 samples have a common behaviour which is well fitted by a unique straight line, the solid line in the graph is a power law dependence $N_S=(t_w/\tau_0)^{0.45T/Tg}$.

Extrapolating this dependence to $T=T_g$ allows a connection with the dynamic scaling hypothesis $\tau \propto \xi^z$ (Eq. 4). Taking $N_S \propto \xi^3$ yields $z=1/0.15=6.7$, in fair agreement with the dynamic scaling fits to the results of *ac* measurements. Recent experiments on a single crystal of the CuMn spin glass very close to $T_g$ ($T/T_g =0.9$) have allowed reaching still larger values of the correlation length, up to 240 lattice units ($\approx$150 nm), almost at the macroscopic scale (Zhai et al, 2019).

Also, experiments on spin-glass thin films have further investigated the relevance of the spin-glass correlation length. When the correlation length reaches the thickness of the sample, modifications of the rejuvenation and memory effects are seen (Ge:Mn layers of 15.5 nm, Guchhait and Orbach, 2015), and a crossover from 3d to 2d behaviour is observed (CuMn layers of 5-20 nm, Zhai et al 2017).

Let us mention that, in experiments exploring the glassy state formed by interacting magnetic nanoparticles, other data in the intermediate time range have been obtained (Nakamae et al, 2012, and references therein). The microscopic flip time of the *superspins* born by the nanoparticles ($10^4$-$10^6$ ferromagnetically coupled spins in each nanoparticle) is much longer than $10^{-12}$ s, ranging up to milliseconds, in such a way that in $\tau_0$ units the explored times are closer to those in simulations. The data from superspins are in continuity with those from spins in both experiments and simulations.

## 5.2 A bridge between experiments and simulations

The numerical simulations of the Edwards-Anderson model (Eq. 1) allow the exploration of the microscopic organization of the spins, which remains inaccessible to the experiments. But an important difficulty is that, due to frustration, the evolution towards equilibrium is very slow, implying time consuming computations. The experiments on real spin glasses are typically exploring the $10^0$-$10^5$ s time range, which in units of the paramagnetic spin flip time $\tau_0 \approx 10^{-12}$s corresponds to $10^{12}$-$10^{17}$ $\tau_0$. Taking $\tau_0 = 1$ Monte-Carlo (MC) step for comparison, the first numerical simulations were exploring up to $\approx 10^7$ MC steps, a rather short-time regime compared with the experiments (Berthier and Young, 2005). In the Janus and Janus II projects, dedicated supercomputers have been built, which allow computation up to $\approx 10^{11}$ MC steps and more.

In simulations, the correlation length has first been determined from a 4-point correlation function (Marinari et al, 1996), which is not directly related to that measured in the above experiments. In a later stage, it has become possible to fully simulate the experimental procedure in which the correlation length is extracted from the effect of the field on the speed of the magnetization relaxation (method used in Fig. 12). Values of the correlation length up to 17 lattice units could be reached (Baity-Jesi et al, 2018). Since the same method can be used in both experiments and simulations, very interesting comparisons can now be made (Zhai et al 2019), recently yielding to common publications by experimentalists and theoreticians (Zhai et al, 2020, Paga et al, 2021). Some discrepancies have been found for large applied fields or close to $T_g$, but a coherent overall description of the results could be reached by defining appropriate scaling laws. The successful modelling of important experimental features of real spin glasses is a crucial progress for a better understanding of complex systems in general.

There has already been an active search for rejuvenation and memory effects in numerical simulations of the Edwards-Anderson model. For Ising spins, they were initially found in d=4, but not in d=3 (Berthier and Bouchaud, 2002, and references therein). For Heisenberg spins, which imply still longer computation times, rejuvenation and memory effects have been seen, but the comparison of the correlation growth in Ising and Heisenberg systems goes at variance between simulations (Berthier and Young, 2005) and experiments (Bert et al, 2004), so the situation is not fully clear. It is likely that the correlation lengths $L_{T-\Delta T}$, $L_T$ (Section 4.3) which could be reached until now by simulations at different temperatures were not spread enough to allow $L_{T-\Delta T} \ll L_T$ and a clear characterization of rejuvenation and memory effects. More space is needed for a full hierarchy of embedded lengths to be at play. Very recently, at the time when this article is being completed, the Janus collaboration could overcome this fundamental hurdle, and succeed in observing rejuvenation and memory effects in the simulated relaxation of the ZFC magnetization (Baity-Jesi et al, 2022).

Temperature chaos is an issue that deserves a new light. It has long been considered as a possible key to rejuvenation (Jönsson et al, 2004), which is a dynamical effect, whereas temperature chaos has been defined in an equilibrium context (Bray and Moore, 1987). Also, it should allow the existence of memory despite rejuvenation. In both Ising and Heisenberg early simulations, the authors indicated that they found no sign of temperature chaos. In the recent large-scale simulations (L=160) by the Janus collaboration, some first signs of a dynamic temperature-chaos effect have been found (Baity-Jesi et al, 2021). In their most recent results, chaos could be clearly characterized, and related with the occurrence of rejuvenation effects (Baity-Jesi et al, 2022).

# 6. Conclusions

Within the wide class of disordered materials, spin glasses occupy a special place because of their conceptually simple definition of randomly interacting spins. Spin glasses are disordered magnetic materials which all share a number of common properties, laying down the definition of a generic spin glass behaviour.

The spin glass state develops below a well-defined temperature $T_g$, above which the spins are in paramagnetic state. At $T_g$, the transition to the spin-glass state presents the same features as for a thermodynamic phase transition to an ordered state, but no macroscopic spin alignment can be seen like in ferromagnets or antiferromagnets. The spin-glass order has a mysterious hidden symmetry, corresponding to spin configurations minimizing the spin-spin interaction energies, which have randomly distributed values.

At $T_g$, a frozen state is established, but this state is not at equilibrium, it is not totally frozen. After cooling from the paramagnetic phase, starting from a random configuration, the spin-glass order is progressively establishing at longer and longer range, causing aging phenomena that show up in the measured magnetic properties. The characteristic size of the growing cooperative regions has been determined in specific experiments, defining a correlation length for the spin-glass order. Recent developments of numerical simulations, taking benefit of powerful custom-built supercomputers, allow very interesting comparisons of this correlation length of a hidden order in real spin glasses and in the conceptually simple (although highly non-trivial) theoretical models (Paga et al, 2021). A better understanding of these models has important consequences, because they are now part of a tool kit of statistical physics which is used for a wide variety of "complex system" problems, such as econophysics, biology, data compressing, optimization etc.

No simple way has been found to bring a real spin glass closer to an equilibrium state, whose nature is still debated (is it unique, or multiple ?). A slower cooling may help partly, as can be efficiently done for structural glasses, but as the temperature is lowered new processes need to be equilibrated, and rejuvenation phenomena show up. Yet, the memory of previous aging states prepared during the cooling can be retrieved upon re-heating. The ability of a spin glass to store several memories of aging at different temperatures is astonishing. A detailed microscopic understanding of the rejuvenation and memory effects at the scale of spin configurations is still lacking. The forthcoming numerical simulations of spin-glass models will probably allow exploring the hierarchy of embedded correlation lengths that is thought to be necessary to observe rejuvenation effects while keeping memory.

In structural and polymer glasses, the dominant scenario is the continuation of aging from one temperature to another. Still, using appropriate procedures, some rejuvenation and memory processes could often be found (as we showed in the case of gelatin). Structural glasses may pertain to a slightly different class of models (spin glass with p-spin interactions, 1-step RSB model, Bouchaud et al, 1996), in which no hierarchy of pure states is found. What are the necessary features of a model in order to obtain rejuvenation and memory phenomena ? This has not yet become fully clear. It is an exciting and important challenge to develop a common understanding of glassy systems, and the spin-glass models clearly have a powerful potential in this direction.

On the side of the experiments on real spin glasses, more and more accurate and refined measurements bring a lot of new information (e.g. Zhai et al, 2022). Anomalous Hall effect measurements bring in principle a direct access to spin-chirality freezing (Kawamura, 1992); if they could be pursued, they should shed some light on the nature of the spin-glass transition for Heisenberg spins (Pureur et al 2004, Taniguchi 2007). However, one may consider that, for making a significant breakthrough in the understanding of real spin glasses, it is necessary to find some way to access to spin configurations. The key may lie in the study of mesoscopic samples (difficult, due to weak signals), in which the growing correlation length may hit the sample size or a "chaos length" (as for the thin layers in Guchhait and Orbach, 2015, Zhai et al, 2017). Transport measurements on mesoscopic wires may offer some insight in the spin configurations by the universal conductance fluctuations (Carpentier and Orignac, 2008), and provide new perspectives on the spin-glass transition (Forestier et al, 2020).

# Acknowledgements

The author deeply thanks V. Dupuis and F. Ladieu for their detailed re-reading of the manuscript and highly relevant suggestions. He also benefitted a lot from very helpful discussions with T. Chakraborty, H. Kawamura, E. Marinari, S. Miyashita and S. Nakamae. H. Bouchiat and A. Parker are warmly thanked for the permission to reuse their figures.